\documentclass[a4paper,11pt]{article}
\pdfoutput=1
\usepackage{jcappub,xcolor}
\usepackage[T1]{fontenc}
\usepackage[shortlabels]{enumitem}
\usepackage{amsmath}
\usepackage{comment}
\usepackage{booktabs}
\usepackage{multirow}
\usepackage{rotating,tabularx}

\usepackage{hyperref}

\begin{document}
\title{Revisiting the Concordance $\Lambda$CDM model using Gamma-Ray Bursts together with Supernovae Ia and Planck data}

\author[a]{Shahnawaz A. Adil,}
\author[b,c,d]{Maria G. Dainotti,}
\author[e]{Anjan A. Sen}

\affiliation[a]{Department of Physics, Jamia Millia Islamia, New Delhi-110025, India}
\affiliation[b]{National Astronomical Observatory of Japan, 2 Chome-21-1 Osawa, Mitaka, Tokyo 181-8588, Japan}
\affiliation[c]{The Graduate University for Advanced Studies (SOKENDAI),
2-21-1 Osawa, Mitaka, Tokyo 181-8588, Japan}
\affiliation[d]{Space Science Institute, Boulder, Colorado}
\affiliation[e]{Centre for Theoretical Physics, Jamia Millia Islamia, New Delhi-110025, India}

\emailAdd{shazadil14@gmail.com}
\emailAdd{maria.dainotti@nao.ac.jp}
\emailAdd{aasen@jmi.ac.in}
\abstract{The Hubble constant, $H_0$, tension is the tension among the local probes, Supernovae Ia, and the Cosmic Microwave Background Radiation. This tension has persisted for decades and continues to puzzle the community. Here, we add intermediate redshift probes, such as Gamma-Ray Bursts (GRB) and Quasars (QS0s), to check if and to what extent these higher redshift probes can reduce this tension. We use the three-dimensional fundamental plane relation among the prompt peak luminosity, the luminosity at the end of the plateau emission, and its rest frame duration. We find similar trend in GRB intrinsic parameters as previously seen in Pantheon-Plus intrinsic parameters. We find an apparent $3.14\sigma$ tension for the GRB intrinsic parameter $b$. Indeed, this tension disappears and the parameters are actually compatible within $2.26\sigma$. Another interesting point is that the 3D relation plays an important role in conjunction with Supernovae data with Pantheon Plus and that this apparent discrepancy shows the  importance of the correction for selection biases and redshift evolution. The incorporation of redshift evolution correction results in a reduction of the GRB tension to $2.26\sigma$ when adjusting correction parameters. We envision that with more data this indication of tension will possibly disappear when the evolutionary parameters of GRBs are computed with increased precision.} %This will highlight the dependence of the intrinsic feature of the energy transfer from the prompt to the afterglow on the cosmological parameters in a non-negligible way. Indeed, the percentage shift of b is--.
\maketitle

\section{Introduction}
\label{sec:introduction}
The standard concordance $\Lambda$CDM model constituting the Dark Energy term as a cosmological constant (CC) $\Lambda$ with the cold dark matter has been a pivotal framework in explaining both the early time observations~\cite{Planck:2018vyg,Alam_2021} as well as late time observations~\cite{SupernovaSearchTeam:1998fmf,SupernovaCosmologyProject:1998vns} providing a solid foundation for understanding the observed acceleration of the Universe. Despite its success, the intrinsic nature of the dark energy component in this model remains a mystery. It is commonly characterized by negative pressure and positive energy density, a framework consistent with the requirements for cosmic acceleration~\cite{Huterer:2017buf}. The $\Lambda$ constant has been a fitting candidate, arising from the vacuum energy density of the Universe rooted in quantum field theory~\cite{Carroll:2000fy}. However, the theoretical construction needs to be revised in accounting for the observed value, introducing a substantial discrepancy of the order of $\mathcal{O}(121)$~\cite{Weinberg:1988cp}. Despite these theoretical challenges, the $\Lambda$CDM model stands as the most favoured choice in accommodating a diverse range of cosmological observations. Nevertheless, recent years have witnessed emerging inconsistencies within the model, particularly in measuring various cosmological parameters from different observational scenarios. Notable examples include tensions affecting the Hubble constant $H_0$~\cite{Verde:2019ivm,Riess:2019qba,DiValentino:2020zio,DiValentino:2021izs,Perivolaropoulos:2021jda,Schoneberg:2021qvd,Shah:2021onj,Abdalla:2022yfr,DiValentino:2022fjm,Hu:2023jqc} and the amplitude of matter fluctuations $S_8$~\cite{DiValentino:2018gcu,DiValentino:2020vvd,Nunes:2021ipq}. These discrepancies suggest the possibility of new physics being at play within the Dark Energy sector, calling for a more comprehensive exploration of alternative models and theoretical frameworks to pursue a more accurate cosmological description.\cite{Eleonora2021,Kamionkowski2022,Sunny2023,Sen:2021wld}\\

One option is to explore the cosmological landscape using intermediate redshift indicators such as Gamma-Ray Bursts (GRBs) and Quasi-Stellar Objects, Quasars (QSOs). Various correlations directly involving the GRB plateau \citep{Dainotti2008, Dainotti2013b, Dainotti2015, Dainotti2016, liang, Bernardini2012a, Xu2012, Margutti, Zaninoni, Shun-Kun2018, Tang2019, Zhao2019, Srinivasaragavan, Dainotti2020a,Wen2020} have been employed as cosmological probes \citep{Cardone, Postnikov, Dainotti2013a, Izzo2015}. For an in-depth discussion on the prompt and prompt-afterglow relations, their associated selection biases, and their application as cosmological tools, refer to \citep{DaiDel2017, DaiAm2018, Dainotti2018}. One notable correlation among these is the Dainotti relation, linking the time at the end of the plateau emission measured in the rest frame, $T^{*}_X$, with the corresponding X-Ray luminosity of the light curve, $L_X$ \citep{Dainotti2008} (see Equation \ref{Dainottiisotropic}). The validity of this correlation is theoretically supported by the magnetar model \citep{Dall'Osso, Bernardini2012b, Rowlinson}. Its three-dimensional extension, incorporating the prompt peak luminosity, $L_{\rm peak}$ \citep{Dainotti2016, Dainotti2017c}, is the fundamental plane correlation or the 3D Dainotti relation.

We use a non-linear relationship between ultraviolet (UV) and X-ray luminosities regarding QSOs. This relationship was initially identified in the early X-ray surveys \citep{1979ApJ...234l...9T,1981ApJ...245..357Z,1986ApJ...305...83A} and has since been validated using diverse QSO samples observed across a broad redshift range and spanning over five orders of magnitudes in UV luminosity with major X-ray observatories \citep[e.g.,][]{steffen06,just07,2010A&A...512A..34L,lr16,2021A&A...655A.109B}. To address the issue of circularity, where the computation of cosmological parameters relies on the assumed underlying cosmological model, we correct for evolutionary effects.
Studies have been conducted in this domain \citep{Dainotti2022QSO} by correcting this relation for selection biases and redshift evolution.
Circular dependencies arise because the determination of cosmological parameters is contingent upon the assumed cosmological model, and parameters derived under such assumptions can bias cosmological results \cite{lenart2023}, mainly when correlations involve luminosities or energies. This study introduces novel aspects compared to previous analyses. Firstly, we incorporate QSOs alongside Supernovae Type Ia (SNe Ia), GRBs, and Cosmic Microwave Background (CMB). Additionally, we apply the RL relation not only in its conventional form, commonly used for cosmological studies, but also consider selection biases and redshift evolution of luminosities through robust statistical methods, thereby overcoming the circularity problem \citep{Dainotti2022QSO}. Secondly, we explore how cosmological results vary with and without corrections for evolution and conduct a comparative analysis of these outcomes.

The subsequent sections of this paper are structured as follows. In Section~\ref{sec:method}, we discuss about the background model and the likelihoods used to investigate various aspects of cosmological parameters, including a discussion on the $M_{B}$ tension. Section~\ref{sec:results} focuses on the implications of the mentioned likelihoods in \ref{subsec:likelihood} on both the model and intrinsic parameters of diverse cosmological observations. Within this section, Subsection~\ref{subec:consistency} discusses consistency analysis, while Subsection~\ref{subec:impact} highlights the effects of different likelihoods such as Planck and SH0ES on GRB intrinsic parameters. Our closing remarks are presented in Section~\ref{sec:conclusions}.

\section{Methodology and Analysis}
\label{sec:method}

\subsection{Background}
\label{subsec:background}
To study the behaviour of the intrinsic parameters of different cosmological observations we shall use the background concordance $\Lambda$CDM model in our analysis. The hubble parameter for $\Lambda$CDM model is given by,
\begin{eqnarray}
H^2(z)=H_0^2 \left [ \Omega_r(1+z)^4+\Omega_m(1+z)^3+\Omega_{\Lambda} \right ] \,,\nonumber \\
\label{eq:hubblerate}
\end{eqnarray}
where $H_{0}$ is the Hubble constant, and $\Omega_{r},\Omega_{m},\Omega_{\Lambda}$ are the present day values for the density parameters of radiation, matter and dark energy respectively. In our work we have considered the Universe to be flat as precisely predicted by Planck experiments \textcolor{red}{\cite{Planck:2018vyg}}, which gives the constraint on the density parameters as follows:
\begin{eqnarray}
\Omega_r+\Omega_m+\Omega_{\Lambda} = 1\,.
\label{eq:flat}
\end{eqnarray}

With this background evolution~\ref{eq:hubblerate} the standard cosmological $\Lambda$CDM model is characterised by 6 parameters, i.e. the baryon and cold dark matter densities $\Omega_{\rm b}h^2=\omega_b$ and $\Omega_{\rm cdm}h^2=\omega_{cdm}$, the optical depth $\tau$, the amplitude and spectral index of the scalar fluctuations $\log[10^{10}A_{s}]$ and $n_s$, and the angular size of the horizon at the last scattering surface $\theta_{\rm{MC}}$.

\subsection{Data }
\label{subsec:likelihood}
In order to obtain the constraint on these mentioned parameters we have used the data detailed below.
\begin{itemize}
\item \textbf{CMB}: From the \textit{Planck} 2018 legacy data release, we use the CMB measurements, namely the high-$\ell$ \texttt{Plik} TT likelihood (in the multipole range $30 \leq \ell \leq 2508$), TE and EE (in the multipole range $30 \leq \ell \leq 1996$),  low-$\ell$ TT-only ($2 \leq \ell \leq 29$), the low-$\ell$ EE-only ($2 \leq \ell \leq 29$) likelihood~\cite{Planck:2019nip}, in addition to the CMB lensing power spectrum measurements \cite{Planck:2018lbu}. We refer to this dataset as \texttt{Planck}.

\item \textbf{Pantheon-Plus}: We use the SNe Ia distance moduli measurements from the Pantheon-Plus sample \cite{Brout_2022}, which consists of 1550 distinct SNe Ia ranging in the redshift interval $z \in [0.001, 2.26]$ and in total has 1701 with duplicated entries. We refer to this dataset as \texttt{Pantheon-Plus}. We also consider the SH0ES Cepheid host distance anchors, which facilitate constraints on both $M$ and $H_0$. When utilizing SH0ES Cepheid host distances, the SNe Ia distance residuals are modified following the relationship eq. (14) of \cite{Brout_2022}. We refer to this dataset as \texttt{Pantheon-Plus\&SH0ES}. 
\item \textbf{GRB}: The Platinum Sample, first defined in \citep{Dainotti2020a}, a subset derived from the Gold Sample defined in \cite{Dainotti2016} has some similarities with analogous samples discussed in the literature \citep{Xu2012,Tang2019}. The Gold Sample is established based on specific criteria for the plateau, where its commencement ($T_t$) must comprise a minimum of five data points, and its slope should be less than $41^{\circ}$— both these conditions were already present in the definition of the Gold sample \citep{Dainotti2016}. The definition of this angle stem from imposing a Gaussian distribution fits for plateau angles and identifying outliers beyond $41^{\circ}$.

The creation of the Platinum Sample involves additional requirements for the plateau: 1) the end time of the plateau ($T_{X}$) should not coincide with observational gaps in the light curves (LCs), ensuring direct determination from data rather than relying on LC fitting; 2) the plateau duration must be at least 500 seconds so that we do not include dubious cases where the plateau is masked out; and 3) no flares should occur at the onset or throughout the entire plateau duration. These additional criteria refines the criteria introduced by \cite{Xu2012}.

Applying these stringent criteria mentioned above, we obtain a Platinum Sample consisting of 50 GRBs out of the 222 X-ray plateaus analyzed. The farthest GRB in this sample is located at redshift $z=5$.
The luminosities both in the prompt and in the plateaus and the times at the end of the plateaus carry error bars which are comparable, namely the $\frac{\Delta_{x}}{x}$, $\frac{\Delta_{y}}{y}$ and $\frac{\Delta_{z}}{z}$ are of the same order of magnitude, where $\Delta_{x}$ is the error on the x-axis (error on time in our case) and $\Delta_{y}$ is the error on y axis (luminosity at the end of the plateau phase) and $\Delta_{z}$ is the uncertainty on the z-axis, which is the peak luminosity. Thus, it is necessary to adopt methods which take into account both error bars like the \cite{Dago05,Kelly2007}.
In addition, this method allows us to introduce an intrinsic scatter, $\sigma_{int}$.
\noindent The fundamental plane relation has the following form:
\begin{equation}
\log L_X = c + a\cdot \log T^{*}_{X} + b \cdot( \log L_{peak}),
\label{Dainottiisotropic}
\end{equation} 
\noindent where $a$ and $b$ are the  parameters given by the \cite{Dago05} which identifies the linear combinations among the features of $\log T^{*}_{X}$ and $\log L_{peak}$, respectively, while $c$ is the normalization. In our analysis we have fixed $c$ to $25.4$ to the best fit parameters for a flat $\Lambda$CDM model. One can argue that fixing c there is a sort of circularity problem, however the uncertainties on this parameter is sufficiently large so that it can include the several cosmological models. Thus, fixing this parameter does not prevent from loosing generality for our fitting procedure.

We use the fundamental plane correlation both with and without the correction computed by the Efron \& Petrosian method (\cite{1992ApJ...399..345E}, hereafter EP) method to see if this correction may carry a reduction on $\sigma_{int}$, and consequently on the cosmological parameters. 
 We derive $\mu_{obs,GRBs}$ in such a way that it is completely independent from the $\mu_{SNe}$, by manipulating the fundamental plane relation corrected for evolution:

\begin{equation}
\begin{split}
\mu_{obs,GRBs}= 5 (b_{1} \log F_{p,cor}+a_{1} \log F_{X,cor} + c_{1}+d_{1} \log T^{*}_X)+25 
\end{split},
\label{equmu}
\end{equation}

\noindent where $\log F_{p,cor}$ and $\log F_{X,cor}$ are the prompt and afterglow emission fluxes, respectively, corrected by the $K$-correction. For the case where we do not take into account the evolutionary effects, considering the relation between fluxes and luminosities we obtain isolating the luminosity distance the following:
\begin{eqnarray}
\begin{split}
 \log_{10}(d_L)&=-\frac{\log_{10} F_{X}+\log_{10}K_{X}}{2 (1-b)}+\frac{b \cdot (\log_{10} F_{peak}+\log_{10}K_{peak})}{2 (1-b)}\\
&\quad -\frac{(1-b)\log_{10}(4\pi)+C}{2 (1-b)}+ \frac{a \log_{10} T^{*}_X}{2 (1-b)}.
 \label{distancemoduli}
 \end{split}
\end{eqnarray}
\noindent where $K_{peak}$ and $K_{X}$ are the $K$-corrections computed for the prompt and the afterglow, respectively, and $a$, $b$, and $c$ are the coefficients of the fundamental plane correlation. We fix in the equation the $C$ parameters and since it carries the largest uncertainties, thus leaving it free would not allow to constrain efficiently the cosmological parameters \cite{Dainotti2023}.

\item \textbf{QSO} The selection of QSOs utilized in this study aligns with the specifications outlined in \citep{2020A&A...642A.150L}, but we here remind that following \cite{Dainotti2022QSO} we use the full sample of QSOs and not a sub-sample so that we can cover the full redshift range. This particular dataset comprises 2421 sources spanning a redshift range extending up to $z = 7.54$ \citep{banados2018}. These QSOs were carefully chosen for their relevance in cosmological investigations, and for a comprehensive account of their selection process, we direct readers to \citep{rl15}, \citep{lr16}, \citep{rl19}, \citep{salvestrini2019}, and \citep{2020A&A...642A.150L}.  As anticipated, the strategy to compute QSO distances makes use of the non-linear relation between their UV and X-ray luminosity \citep{steffen06,just07,2010A&A...512A..34L,lr16,2021A&A...655A.109B}, namely 
\begin{equation}\label{RL}
\mathrm{log} L_{X} = g \, \mathrm{log} L_{UV} + b
\end{equation}
where $L_X$ and $L_{UV}$ are the luminosities (in $\mathrm{erg \, s^{-1} \, Hz^{-1}}$) at 2 keV and 2500 \AA, respectively. In \cite{DainottiQSO} this relation has been corrected for selection biases and redshift evolution using the EP and it has been proved that it is intrinsic to QSO's properties so that it can be reliably used to standardize QSOs as cosmological tools.
To make use of Equation \eqref{RL}, we compute luminosities from measured flux densities $F$ according to $L_{X,UV}= 4 \pi d_{L}^{2}\ F_{X, UV}$, where $d_L$ is computed under the assumption of a cosmological model using the corresponding parameters as free parameters of the fit. Usually, luminosities of the sources have to be corrected for the K-correction, defined as $1/(1+z)^{1-\beta}$, where $\beta$ is the spectral index of the sources when we assume a simple power law for the spectrum, but for QSOs it is assumed to be $\beta = 1$, leading to a $K=1$. So the K-correction has been omitted following \cite{2020A&A...642A.150L}.

\end{itemize}

\subsection{Methodology} 
\label{subsec:methods}
We have performed the statistical analysis using the Bayesian inference technique, which is extensively used for parameter estimation in cosmological models. Bayesian inference provides a powerful framework for updating the probability of a hypothesis as more evidence or information becomes available. In this context, the posterior probability distribution function of the model parameters is determined by combining the likelihood function with the prior probability of the model parameters. Mathematically, the posterior distribution is proportional to the product of the likelihood function and the prior distribution. To estimate the model parameters, we commonly use multivariate Gaussian likelihoods. These likelihoods quantify how well the model parameters explain the observed data and are given by the expression:
\begin{equation}
{\cal L} (\Theta) \propto \exp\left[-\frac{\chi^2(\Theta)}{2}\right],
\end{equation}
where $\Theta$ represents the set of model parameters. Here, $\chi^2(\Theta)$ is the chi-squared statistic, which measures the difference between the observed data and the values predicted by the model for a given set of parameters. In our analysis, we have adopted a uniform prior distribution for the model parameters. A uniform prior indicates that all parameter values are initially considered equally likely before observing the data. Under this assumption, the posterior probability distribution simplifies to being directly proportional to $\exp\left[-\frac{\chi^2(\Theta)}{2}\right]$. Therefore, minimizing the chi-squared statistic $\chi^2(\Theta)$ maximizes the likelihood function and, consequently, the posterior probability. This means that the parameter set $\Theta$ that results in the smallest $\chi^2(\Theta)$ is the most probable given the observed data. The likelihoods for all corresponding data sets are defined in a similar manner, ensuring that we can comprehensively evaluate the goodness-of-fit for different parameter sets across various observations. We have defined the different likelihoods as the following:

\begin{itemize}
  \item   \textbf{Planck:} The Planck likelihood is associated with the CMB temperature and polarization spectra. It is defined in two multipole ranges: $\ell \leq 30$ and $\ell \geq 30$. For high-$\ell$ values, the likelihood has two components: 1) the baseline Plik likelihood and 2) the CamSpec likelihood. For low-$\ell$ values, it consists of the low-$\ell$ temperature Commander likelihood and the low-$\ell$ SimAll polarization likelihood. Additionally, we have utilized the lensing likelihood of the CMB spectrum. We used the publicly available Planck likelihood code\footnote{\href{https://pla.esac.esa.int/\#cosmology}{Planck Likelihood code}} to calculate these likelihoods. Detailed implementation can be found in Section 2.2 in \cite{Planck:2018vyg}; see also \cite{Planck:2019nip}. 
 \item \textbf{Pantheon-Plus:}The likelihood of Pantheon-Plus is based on the measurement of distance modulus ($\mu$), which is given by Tripp's formula \cite{BBC}  as:
\begin{equation}
    \mu=m_B + \alpha x_1 - \beta c - M  - \delta_{\rm bias} +\delta_{\rm host},~\label{eq:Tripp}
\end{equation}
 where $\alpha$ and $\beta$ are stretch and color (respectively) parameters, $M$ is the absolute magnitude of an SN~Ia, calibrated by Cepheids, $\delta_{\rm bias}$ is a correction term.
One can constrain cosmological models with SN data using $\chi^2$ mimimalization (as done in \cite{riess98,Astier06}) and including the  systematic covariance in \cite{Conley11}. We have used the similar formalism of \cite{Conley11} where  $\chi^2$ likelihood is defined as:
\begin{equation}
\label{eq:likelihood}
%-chisq - ln(det(cov)) / 2
-2{\rm ln}(\mathcal{L}(SN)) = \chi^2 = \Delta \vec{D}^T~C_{\rm stat+syst}^{-1}~\Delta \vec{D} ,
\end{equation}
where $\vec{D}$ is the vector of 1701 SN distance modulus residuals computed as
\begin{equation}
\label{eq:dmu}\Delta D_i = \mu_i - \mu_{{\rm model}}(z_i) ,
\end{equation}
and each SN distance ($\mu_i$) is compared to the predicted model distance given the measured SN/host redshift ($\mu_{{\rm model}}(z_i)$). The model distances are defined as 
\begin{equation}
\mu_{{\rm model}}(z_i) = 5\log(d_L(z_i)/10\,{\rm pc}) ,
\end{equation} 
where $d_L$ is the luminosity distance which is defined using the expansion history $H(z)$. For a flat $\Lambda$CDM model, the luminosity distance is described by
\begin{equation}
\label{eq:dl}
d_L(z) = (1+z)c\int_0^{z}\frac{dz^\prime}{H(z^\prime)},
\end{equation}
where $d_L(z)$ is calculated at each step of the cosmological fitting process, and the parameterization of the expansion history (used in Eq.~\ref{eq:dl} and therefore in the likelihood in Eq.~\ref{eq:likelihood}) in this work, where $H(z)$ is defined in Eq.~\ref{eq:hubblerate}

The parameters $M$ (Eq.~\ref{eq:Tripp}) and H$_0$ (Eq.~\ref{eq:hubblerate}) are degenerate when analyzing SNe alone.

When using SH0ES Cepheid host distances, the SN distance residuals are modified as:

\begin{equation}
\label{eq:dmuprime}
\Delta D^\prime_i=
        \begin{cases}
            \mu_i - \mu_i^{{\rm Cepheid}} & i \in \text{Cepheid hosts} \\
            \mu_i - \mu_{{\rm model}}(z_i) &\text{otherwise} ,
        \end{cases}
    \end{equation}
where $\mu_i^{{\rm Cepheid}}$ is the Cepheid calibrated host-galaxy distance provided by SH0ES and where $\mu_i - \mu_i^{{\rm Cepheid}}$ is sensitive to the parameters $M$ and H$_0$ and is largely insensitive to $\Omega_m$\cite{reiss2021}.  We also include the SH0ES Cepheid host-distance covariance matrix ($C^{\rm Cepheid}_{\rm stat+syst}$) presented by R22 such that the likelihood becomes
\begin{equation}
\label{eq:sh0eslikelihood}
-2{\rm ln}(\mathcal{L}(SN)^\prime) = \Delta \vec{D^\prime}^T~(C^{\rm SN}_{\rm stat+syst}+C_{\rm stat+syst}^{\rm Cepheid})^{-1}~\Delta \vec{D^\prime} ,
\end{equation}
where $C^{\rm SN}_{\rm stat+syst}$ denotes the SN covariance.

\item \textbf{GRB:} To obtain the likelihood for GRB, We compare both $\mu_{obs, GRBs}$ given in Equ. \ref{equmu} with the theoretical $\mu_{th}$, defined as:

\begin{equation}
\mu_{th}=5\cdot \log\ d_L(z) +25,
\label{modulus}
\end{equation}
 where $d_L(z)$ is defined in the equation~\ref{eq:dl}.

By combining the definitions in Equ. \ref{Dainottiisotropic} and Equ. \ref{distancemoduli} we finally reproduce Equation \ref{equmu}. We then write the likelihood as: 

\begin{align}
\begin{split}
 \mathcal{L}(GRBs) =&
 \sum_{i}\biggl[\log \biggl( \frac{1}{\sqrt{2\pi}\sigma_{\mu,i}} \biggr)- \frac{1}{2}\biggl( \frac{\mu_{th,GRB,i}-\mu_{obs,GRB,i}}{\sigma_{\mu,i}}\biggr)^{2} \biggr] 
 \label{likelihood} 
\end{split}
\end{align}

\item \textbf{QSO:} The likelihood used for QSOs used in our analysis is given as \citep[see also][]{2021MNRAS.502.6140K,2022MNRAS.510.2753K,2022MNRAS.515.1795B,2022arXiv220310558C}:
\begin{equation} \label{lfqso}
ln(\mathcal{L}(QSO)) = -\frac{1}{2} \sum_{i=1}^{N} \left[ \frac{(y_{i}-\phi_{i})^{2}}{s^{2}_{i}} + \text{ln}(s^{2}_{i})\right].
\end{equation}
In this case, the data $y_{i}$ correspond to $\text{log}L_{X}$, while $\phi_{i}$ to the logarithmic X-ray luminosity predicted by the X-UV relation. Moreover, $s^{2}_{i} = \sigma ^{2}_{y_{i}} + g^{2} \sigma ^{2}_{x_{i}} + sv^{2}$ and it takes into account the statistical uncertainties on $\text{log}L_{X}$ ($y$) and $\text{log}L_{UV}$ ($x$), but also the intrinsic dispersion $sv$ of the X-UV relation, which is another free parameter of the fit, as defined in the \cite{Dago05} method. Practically, the likelihood of QSOs is just the same likelihood function used for SNe Ia, but modified to include the contribution of the intrinsic dispersion of the X-UV relation.

\end{itemize}

Along with the 6 standard parameters in baseline $\Lambda$CDM model we have $1$ intrinsic parameter for Pantheon-Plus i.e. $M_{B}$, 3 intrinsic parameters for GRB i.e. $a,b,svGRB$ and 3 intrinsic parameters for QSO i.e. $g,be,svQ$. We have used publically available \texttt{CLASS} \cite{Blas_2011} code to calculated the background parameters. To implement the likelihoods we have used \texttt{MontePython}\cite{Audren:2012wb,Brinckmann:2018cvx} which uses Metropolis-Hastings algorithm to maximize the likelihoods and derive constraints on cosmological parameters as well as the intrinsic parameters. All of our runs were set up to reach a Gelman-Rubin convergence criterion of $R - 1 < 10^{-2}$.

\section{Results}
\label{sec:results}
In this section we shall present the results obtained in our analysis. We have divided this section in 3 subsequent subsections. In the Subsec.~\ref{subec:consistency} we have performed a consistency check of GRB and QSO observations respectively, using the Planck priors. In Subsec.~\ref{subec:impact} we then study the constraints on GRB parameters using Planck and \texttt{Pantheon-Plus\&SH0ES} data. Lastly, in Subsec.~\ref{subec:correction} we have shown the implications of redshift evolution corrections in GRB observations.

\subsection{$M$ Tension}
\label{subsec:Mb}

\begin{table}[!htbp]
    \centering
    \begin{tabular}{cccc}
    \hline
       Observations  & $H_0$(Km/s/Mpc) & $\Omega_m$&References \\
       \hline
      Planck & $67.4\pm0.5$ & $0.315\pm0.007$ & \cite{Planck:2018vyg}\\
      \hline
        SH0ES & $73.30\pm1.04$ & $0.3$ & \cite{reiss2021}\\ 
        \hline
        Pantheon-Plus\&SH0ES  & $73.6\pm1.1$ & $0.334\pm0.018$ & \cite{Brout_2022}\\
        \hline
    \end{tabular}
    \caption{Baseline values of Cosmological parameters in different observations}
    \label{tab:baseline}
\end{table}

\begin{figure}[htbp]
\centering
\includegraphics[scale=0.68]{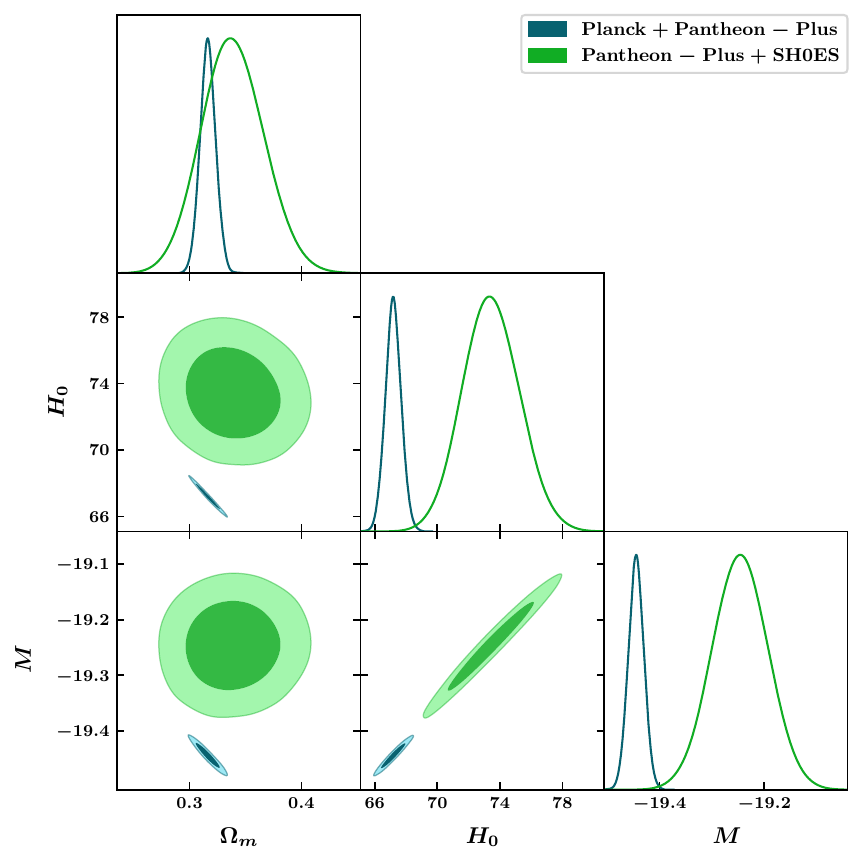}
\caption{Contour plots of the values of $M$ and $H_0$ using the Planck+ Pantheon-Plus data and Pantheon-Plus combined with SH0ES are shown in dark green and light green, respectively.}
\label{fig:mb}
\end{figure}

In our initial analysis utilizing Pantheon-Plus data with Planck or Cepheid priors, we observe a notable correlation between the intrinsic parameter $M$ and the Hubble constant $H_0,$ as depicted in Fig.~\ref{fig:mb}. Using the Planck priors to Pantheon-Plus data yields an $H_0$ value of $67.41\pm0.54,$ consistent with the value derived from Planck priors alone~\cite{Planck:2018vyg}. Conversely, using Cepheid priors on Pantheon-Plus data results in an $H_0$ value of $73.04\pm1.04,$ aligning with the baseline Cepheid analysis by \cite{reiss2021}. Similar trends are observed for other cosmological parameters, as detailed in the reference table 1. Our preliminary findings suggest that Pantheon-Plus data does not significantly alter cosmological parameters, a conclusion in line with recent analyses by the Pantheon-Plus team~\cite{Brout_2022}. We have presented the baseline values for parameters in different observations in table \ref{tab:baseline} for comparison. Notably, the strong influence of $H_0$ on the measurement of $M$ leads us to identify discrepancies in the $M$ measurements~\cite{Camarena2023}, significantly attributed to the well-known Hubble tension.
This discrepancy about the value of $M$ can be seen also as an evolving trend of $H_0$ as already studied previously in \citep{Dainotti2021H0constant,Dainottigalaxies2022}.

\subsection{Consistency test with Planck}
\label{subec:consistency}
\begin{figure*}[htbp]
\centering
\includegraphics[scale=0.50]{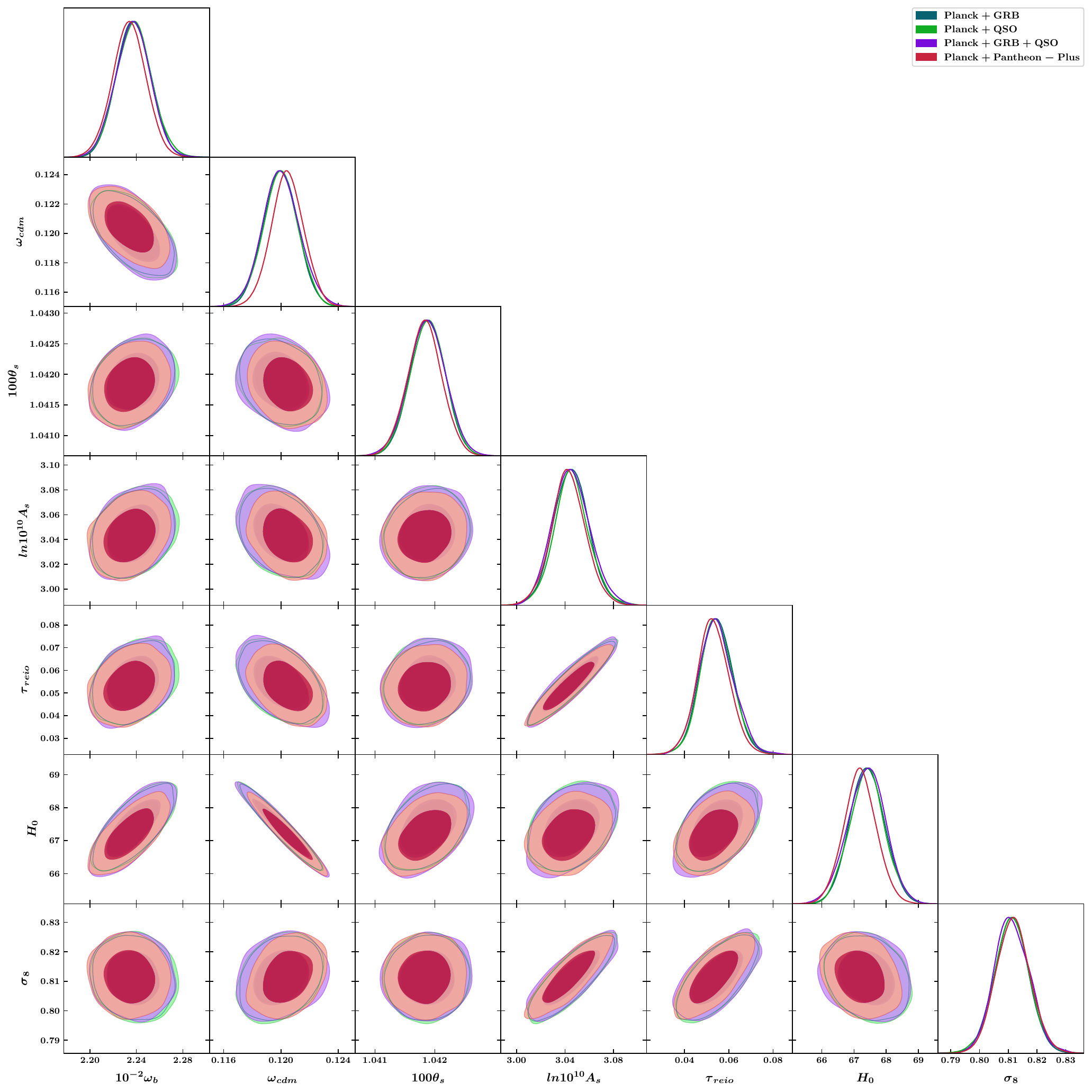}
\caption{The contours plots for Planck+GRBs, Planck +QSOs, Planck+GRBs+QSOs and Planck+ Pantheon-Plus are shown in dark blue, purple and red, respectively.}
\label{fig:all}
\end{figure*}

Firstly, we examine the effect of GRB and QSO data using the Planck priors on different cosmological parameters of standard $\Lambda$CDM model. In Fig. \ref{fig:all} we see that there is no significant statistical difference in the measurement of the cosmological parameters using the GRB and QSO likelihoods as was the case for Pantheon-Plus~\ref{subsec:Mb}. It could be attributed to the strong constraints of Planck likelihood. We have presented the constraints on cosmological parameters in Table~\ref{tab:ii}. It shows that the GRB and QSO data are consistent with Planck.

%%%%%%%%Table for all data set cosmological parameters%%%%
\begin{table}[htbp]
\centering
\begin{tabular} { l  c c c c}

 Parameter &  Planck+GRB & Planck+QSO & Planck+GRB+QSO &Planck+Pantheon-Plus \\
\hline
{\boldmath$10^{-2}\omega{}_{b }$} & $2.238\pm 0.015$ & $2.238\pm 0.015$ & $2.238\pm 0.015$ & $2.238\pm 0.015$\\

{\boldmath$\omega{}_{cdm }$} & $0.1200\pm 0.0012$&$0.1200\pm 0.0012$&$0.1200\pm 0.0012$&$0.1200\pm 0.0012$\\

{\boldmath$100\theta{}_{s }$} & $1.04187\pm 0.00029$&$1.04187\pm 0.00029$&$1.04187\pm 0.00029$&$1.04187\pm 0.00029$\\

{\boldmath$ln10^{10}A_{s }$} & $3.045\pm 0.015$&$3.045\pm 0.015$&$3.045\pm 0.015$&$3.045\pm 0.015$\\

{\boldmath$n_{s }         $} & $0.9653\pm 0.0042$&$0.9653\pm 0.0042$&$0.9653\pm 0.0042$&$0.9653\pm 0.0042$\\

{\boldmath$\tau{}_{reio } $} & $0.0546\pm 0.0075$&$0.0546\pm 0.0075$&$0.0546\pm 0.0075$&$0.0546\pm 0.0075$\\

{\boldmath$H_0             $} & $67.41\pm 0.54$&$67.41\pm 0.54$&$67.41\pm 0.54$&$67.41\pm 0.54$\\

{\boldmath$\sigma_8         $} & $0.8112\pm 0.0060$&$0.8112\pm 0.0060$&$0.8112\pm 0.0060$&$0.8112\pm 0.0060$\\
\hline
\end{tabular}

\caption{All cosmological parameters including the combinations of probes, such as the Planck+GRBs, the Planck+QSOs, Planck +QSOs+GRBs, Planck+Pantheon-Plus from the first to the last column.\label{tab:ii}}
\end{table}
%%%% 

\subsection{Impact on GRB intrinsic parameters}
\label{subec:impact}
\begin{figure}[htbp]
\centering
\includegraphics[scale=0.67]{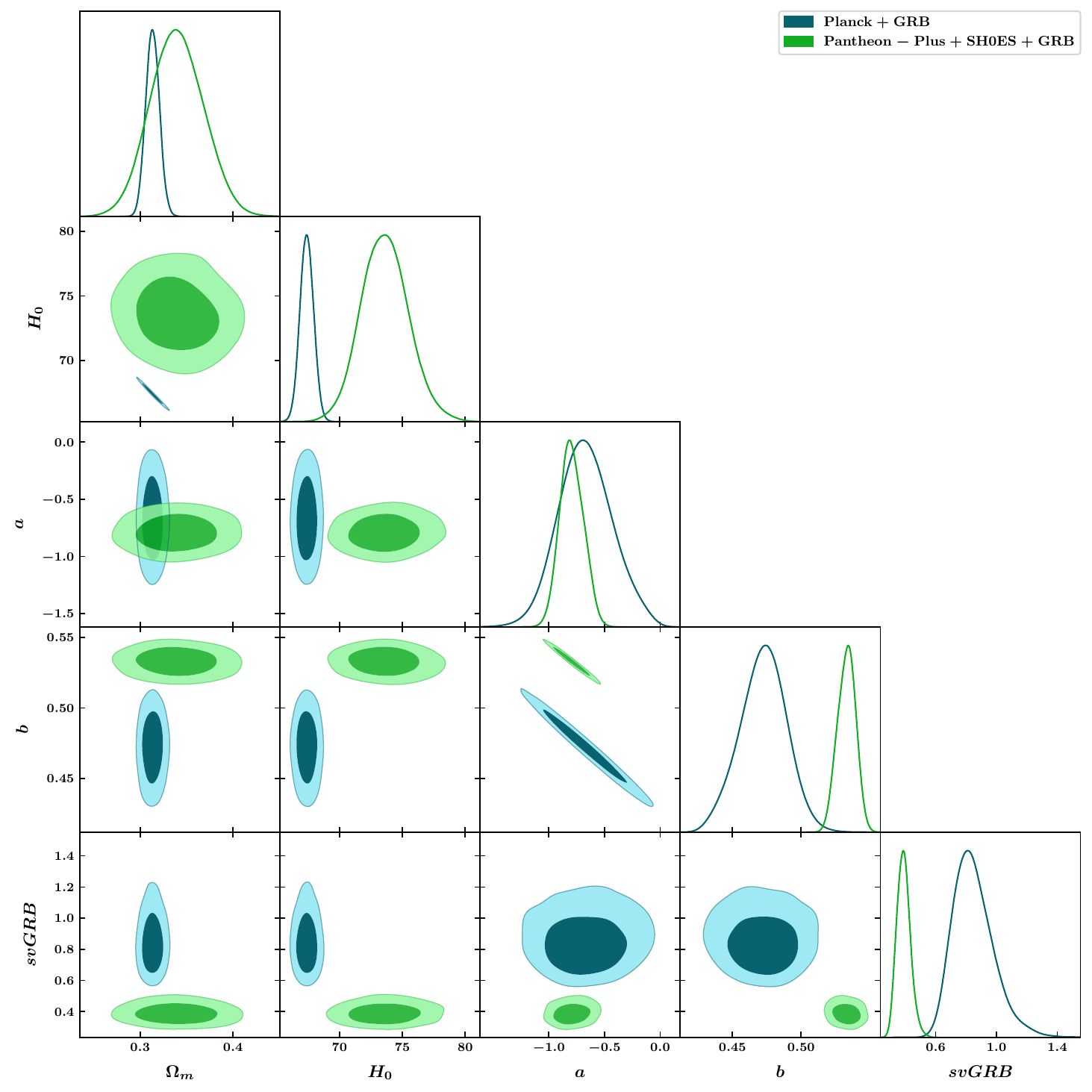}
\caption{The contour plots for the parameters $a$, $b$, $\sigma_v$, and $H_0$ for Planck combined with GRBs and for GRBs combined with Pantheon and with SH0ES data are shown in darker blue and lighter green, respectively.}
\label{fig:grb1}
\end{figure}

Further, we have performed detailed analysis of the impact of Planck and \texttt{Pantheon-Plus\&SH0ES} data on GRB intrinsic parameters. This observation is illustrated in Fig.~\ref{fig:grb1}, where the well-known $H_0$ discrepancy persists with a difference exceeding $2.89\sigma$. While our analysis did not reveal a prominent influence of GRBs inclusion in relation to the cosmological parameters, as depicted in Fig.~\ref{fig:grb1}, we can note some apparent discrepancies in cosmological parameters that seems to have affected the values of GRB intrinsic parameters. This difference is not present in the case of intrinsic parameter $a$  since it fall within the $1\sigma$ range. However, for the parameter $b$ the apparent difference is $3.14\sigma$ and for $svGRB$, the difference is $2.8\sigma$. This apparent difference is due to selection bias and redshift evolution that have not been taken into account yet. 

\begin{figure}[htbp]
\centering
\includegraphics[scale=0.67]{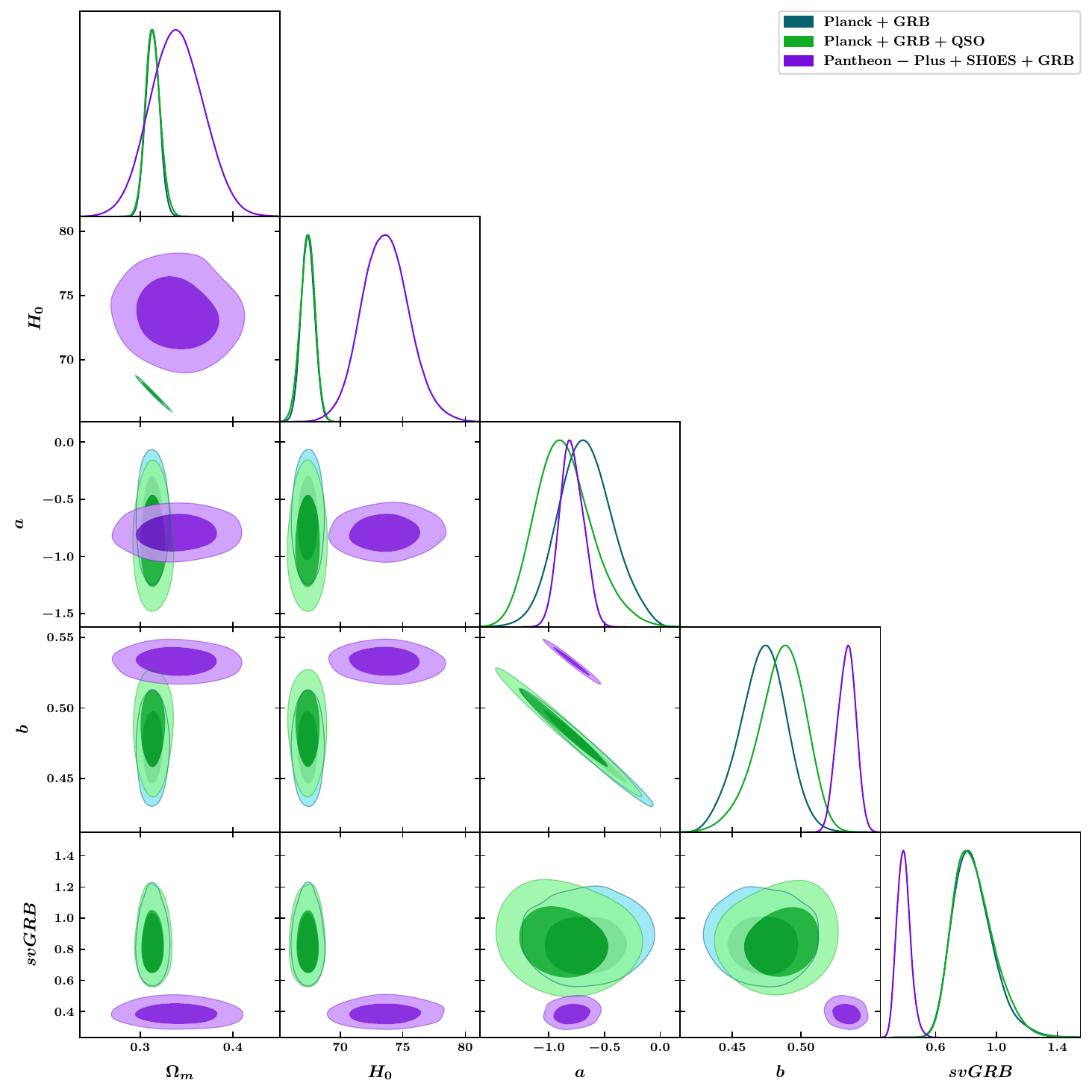}
\caption{The contour plots for different parameters with Planck+GRB in dark blue, Planck+GRB+QSO in green and GRB+Pantheon-Plus\&SH0ES combinations in purple colors respectively.}
\label{fig:grb2}
\end{figure}

In Fig.~\ref{fig:grb2}, we present the impact of incorporating the QSO likelihood into the previous scenarios (see Fig. \ref{fig:grb1}). The contours in Fig.~\ref{fig:grb2} reveals that the QSO likelihood does not influence the GRB parameters $svGRB$ and $a$, but it does show a slight improvement in the value of parameter $b$. The inclusion of the QSO likelihood, represented by the green contours in Fig.~\ref{fig:grb2}, has reduced the discrepancies for $b$ which is  $3.04\sigma$. This analysis, however, does not include the selection effects yet.

\begin{table}[!htbp]
    \centering
    \scalebox{0.6}{
    \begin{tabular}{p{35mm}|lll|lll|lll}
    \toprule[1.2pt]
    \toprule[1.2pt]
    Parameter & \multicolumn{3}{c}{Without evolution} & \multicolumn{3}{c}{With evolution} & \multicolumn{3}{c}{Kvarying}\\
    \cmidrule(lr){2-4} \cmidrule(lr){5-7} \cmidrule(lr){8-10}
     & PGRB  & PSGRB & Tension & PGRB & PSGRB & Tension & PGRB & PSGRB & Tension\\
    \midrule
    $\boldsymbol{a}$ & $-0.67^{+0.23}_{-0.26}$ & $-0.84^{+0.16}_{-0.12}$ &$0.55\sigma$ &$-0.84^{+0.16}_{-0.12}$ &$-0.95\pm0.19$&$0.5\sigma$ & $-0.81_{-0.18}^{+0.14}$& $-0.75 \pm 0.11$&$0.47\sigma$\\ 
    $\boldsymbol{b}$ & $0.473^{+0.018}_{-0.016}$ & $0.5358^{+0.0077}_{-0.0096}$ & $3.14\sigma$ &$0.423\pm 0.017$&$0.4771^{+0.0094}_{-0.0079}$&$3\sigma$&$0.434^{+0.018}_{-0.014}$&$0.4789 \pm 0.0083$&$2.26\sigma$\\
    $\boldsymbol{svGRB}$ & $0.849^{+0.098}_{-0.16}$ & $0.384\pm 0.047$& $2.8\sigma$&$0.572^{+0.073}_{-0.097}$&$0.162^{+0.083}_{-0.097}$&$3.22\sigma$ & $0.391^{+0.087}_{-0.12}$& $< 0.144$ & $2.05\sigma$\\
    $\boldsymbol{H_0}$ & $67.41\pm 0.54$ &$73.4^{+1.7}_{-2.0}$ &$2.89\sigma$& $67.39\pm 0.55$&$73.5\pm 1.8$&$3.25\sigma$&$67.37 \pm 0.55$& $73.5 \pm 1.8$ &$3.26\sigma$\\
    $\boldsymbol{\Omega_m}$ &$0.3134\pm0.0073$ &$0.339^{+0.028}_{-0.033}$ & $0.77\sigma$&$0.3136\pm0.0076$&$0.331^{+0.026}_{-0.030}$&$0.58\sigma$&$0.3139 \pm 0.0075$ & $0.338 \pm 0.029$ & $0.86\sigma$\\
    \bottomrule
    \end{tabular} }
    \caption{$1\sigma$ Constraints on parameters for Planck, GRB, and Pantheon-Plus\&SH0ES data combinations with, without evolution corrections and evolution corrections with Gaussian K corrections along with tensions in the parameters from PGRB (Planck+GRB) and PSGRB (Pantheon-Plus\&SH0ES+GRB) data.}
    \label{tab:all_tab} 
\end{table}

\subsection{Redshift Evolution correction}
\label{subec:correction}
In the study of high-redshift observations of GRBs compared to the Pantheon-Plus dataset limited to redshifts up to $z=2.26$, a recent study (~\cite{Dainotti2023}) has emphasized the importance of considering redshift-dependent evolution for a more nuanced understanding of these astrophysical phenomena. GRB observations in our sample extend to high redshifts, reaching up to $z=5$ and due to large redshift redshifts spanned they necessitate corrections to account for the redshift evolution of their intrinsic parameters.

The redshift evolution correction is incorporated into the analysis through a fundamental plane relation describing the luminosities of GRBs. The relation is expressed as follows:

\begin{eqnarray}
\begin{split}
\log L_X-k_{L_a} \log (z+1) &= a_{e v} \cdot (\log T_X^*-k_{T_X} \log (z+1)) \\
&\quad + b_{e v} \cdot (\log L_{\text{peak}}-k_{L_{\text{peak}}} \log (z+1)) + c_{e v}
\end{split}
\label{eq:correction}
\end{eqnarray}

Here, $a_{\mathrm{ev}}, b_{\mathrm{ev}}$, and $c_{\mathrm{ev}}$ represent parameters incorporating the redshift evolution.
These parameters have been discussed extensively in a series of papers such as \citep{Dainotti2013a,Dainotti2013b,Dainotti2015,Dainotti2017c,Dainotti2020a,Dainotti2023}.
The coefficients $k_{L_{\text {peak }}}, k_{T_X}$, and $k_{L_a}$ are evolutionary coefficients,  with fixed mean values set at $k_{L_{\text {peak }}}=2.24, k_{T_X}=-1.25, k_{L_a}=2.42$ in this analysis, as obtained from (~\cite{Dainotti2023}).

\begin{figure*}[htbp]
\includegraphics[scale=0.37]{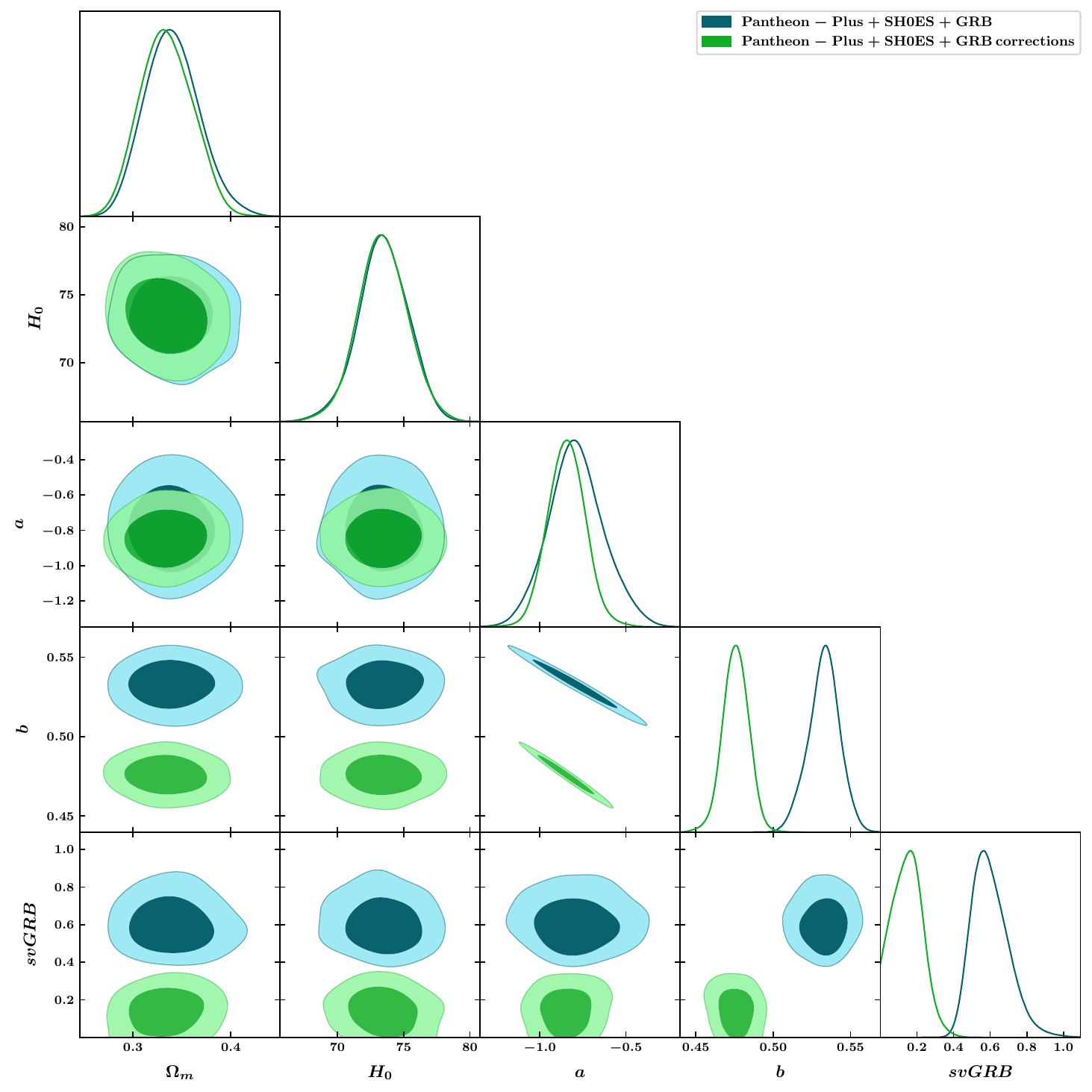}
\includegraphics[scale=0.37]{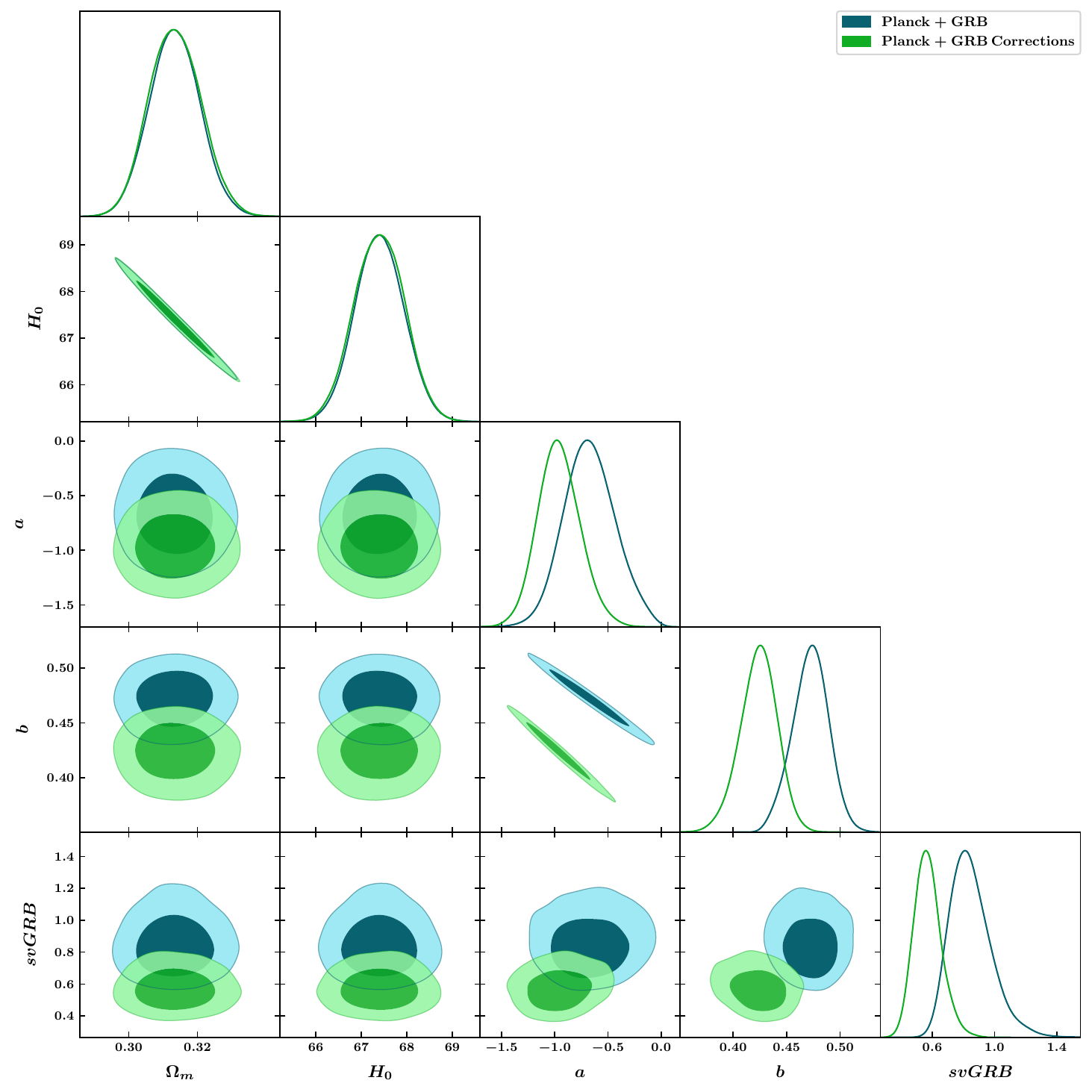}
\caption{Contours showing the effect of evolution correction in Pantheon-Plus\&SH0ES prior in left panel and Planck prior in right panel}
\label{fig:grb3}
\end{figure*}

\begin{figure*}[htbp]
\centering
\includegraphics[scale=0.67]{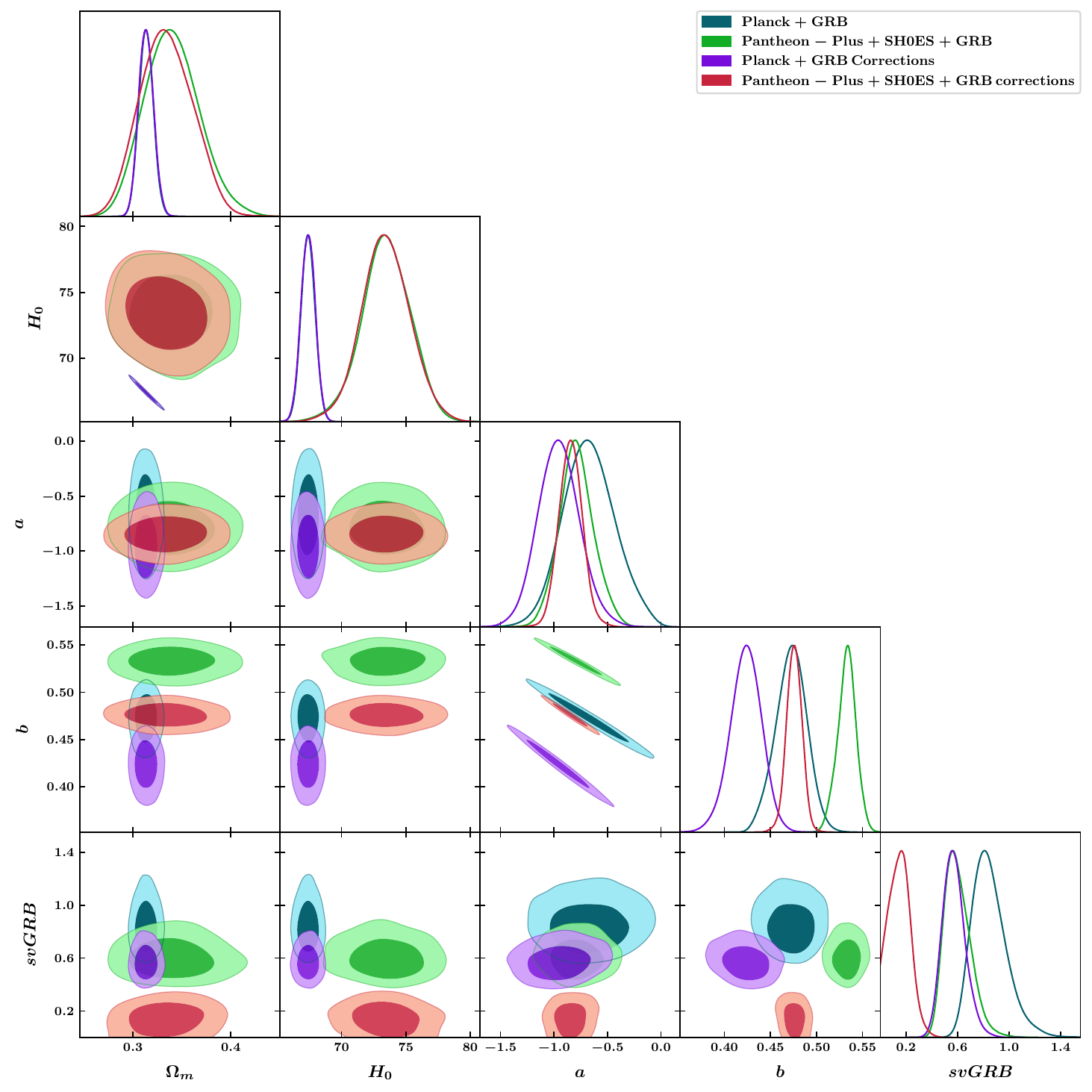}
\caption{Contours showing the effect of evolution correction in for both Pantheon-Plus\&SH0ES and Planck prior combined.}
\label{fig:grb4}
\end{figure*}

The impact of these corrections on the intrinsic parameters of GRBs is visually apparent in Fig.~\ref{fig:grb3}. Notably, the cosmological parameters remain unaffected. The green contours in Fig.~\ref{fig:grb3} illustrate that the values of parameter $b$ and $svGRB$ experience a significant decrease due to the corrections. The extent of decrease is more pronounced when considering the SH0ES prior, but the change is less prominent with Planck priors. Refer Table \ref{tab:all_tab} for shift in the values. Moreover, the corrections induce shifts in the parameter values and also influence the contours of these parameters, as depicted in Fig.~\ref{fig:grb3}. This shows the importance of accounting for redshift evolution to achieve a more accurate understanding of the intrinsic properties of GRBs. Since we are in an era in which we strive for precision cosmology even if the impact on cosmological parameters with the inclusion of GRBs is comparatively minimal, this is still crucial, given the extension up to $z=5$. We present the constraints on the parameters in the Table \ref{tab:all_tab}. Lastly, in Fig.~\ref{fig:grb4} we show the overall effect of evolution correction in GRB parameters. The blue and green contours represents the results without the correction and red and violet colour contour represents the results considering the redshift correction. The contours in red and violet demonstrate a more refined and aligned set of results compared to the blue and green contours, indicating that accounting for the changes over redshift corrections has contributed to a more accurate representation of GRB parameters.

\begin{figure*}[htbp]
\centering
\includegraphics[scale=0.67]{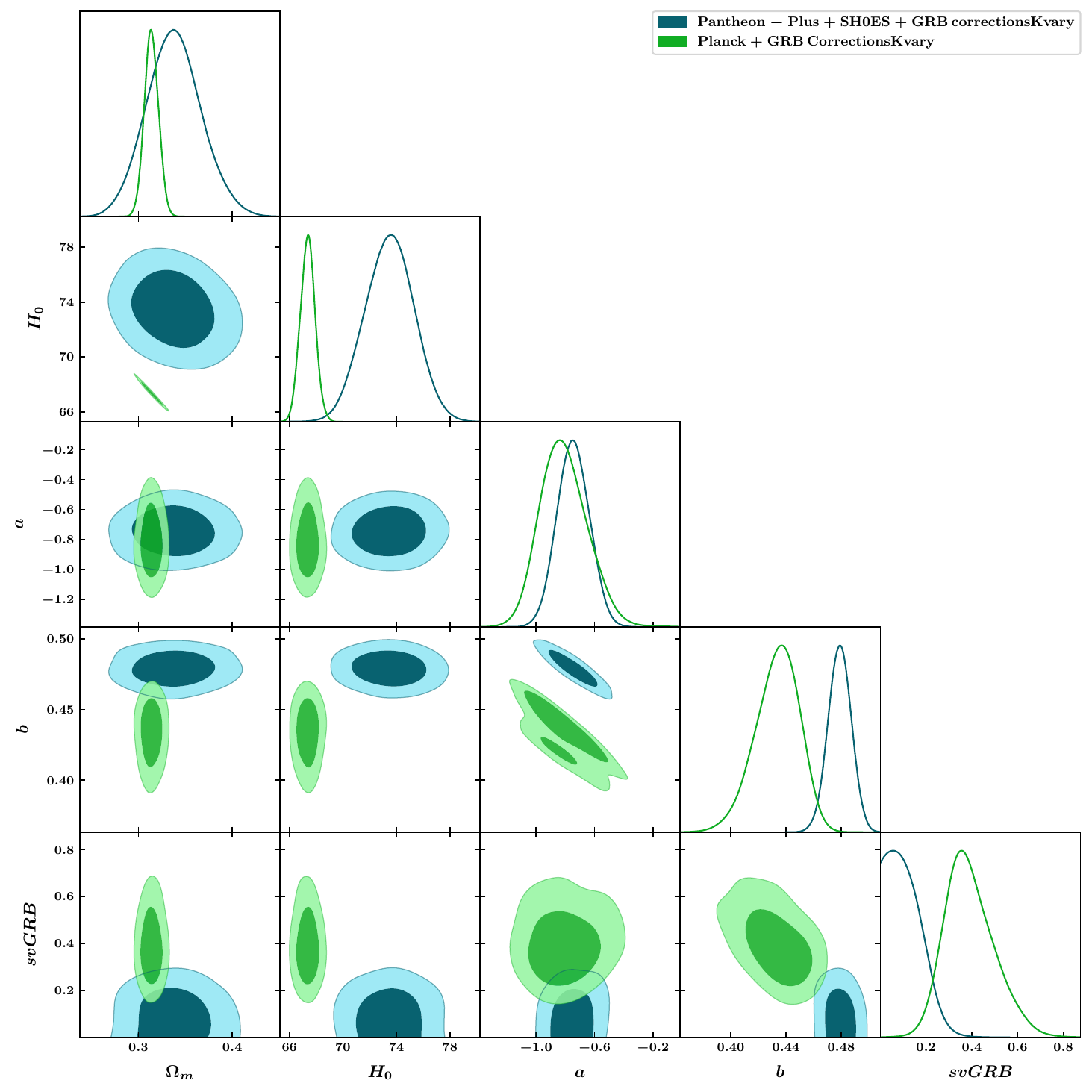}
\caption{Contours showing the effect of evolution correction in for both Pantheon-Plus\&SH0ES and Planck priors when the K corrections  are free}
\label{fig:grb5}
\end{figure*}

We expanded our analysis by allowing the evolutionary correction parameters, namely $k_{L_{\text{peak}}}$, $k_{T_X}$, and $k_{L_a}$ in equation \ref{eq:correction}, to vary freely, aiming for a more realistic estimation. We set Gaussian priors on these evolution parameters. Our findings indicate a notable improvement in results when these evolutionary parameters are left free compared to when they are fixed. Specifically, we find the value of $b$ parameter as $0.48\pm 0.008$ for Pantheon-Plus\&SH0ES prior, while under the influence of Planck prior it is $0.443 \pm 0.018$. This reduction in the apparent tension in the $b$ parameter narrows down to $2.26\sigma$ when the evolutionary parameters are allowed to vary freely which is shown in figure \ref{fig:grb5}. This shows the importance of implementing redshift evolution correction in GRB observations. It is also important to note that the tension in $H_0$A similar trend is evident in the intrinsic parameters of the Pantheon sample, where the absolute magnitude ($M$) shows a shift at higher redshifts \cite{2023MNRAS.520.5110P,Dainotti2021H0constant,Dainottigalaxies2022}. However, it's worth noting that our constraint on the evolution parameters was not as stringent, given that we only considered the 50 platinum sample of GRBs. Future observations are expected to provide a higher number of precise GRB data \cite{2022MNRAS.514.1828D}, which should offer improved constraints and alleviate this apparent tension in the $b$ parameter.

It is important to note that, there is  a reduction of the discrepancies in the $b$ parameter when we compare the case without evolution to the case of the varying evolution. These discrepancies are reduced from 3.14 $\sigma$ to 2.26 $\sigma$, namely it reduces 38\% in $b$. Despite the corrections, some small disparities in the measurements still persist. %This shows the complexity inherent in studying GRBs and suggests that, while corrections improve the overall agreement, there are likely additional factors influencing these cosmic phenomena that are not fully captured by the applied corrections. 
Thus, to further investigate these discrepancies, we also checked if the shift in the $b$ parameter is due to the dependence on the $H_0$, but we did not find a clear trend, so additional investigation is needed. In particular what we observe is that while using only GRB data together with either Planck or SH0ES prior for $H_{0}$, there is no shift in the constraint value for $b$. Taking the full Planck likelihood together with GRB data also results in same value for $b$ showing that Planck likelihood has no effect on the GRB parameters. This is expected as early Universe observations from Planck has minimal effect at low redshift observations like GRBs. But when one adds Pantheon-plus data to the GRB data, the constraint on GRB parameter like $b$ gets shifted due to the interplay between SNe Ia and GRB data at low redshifts. So to summarise, as seen in figure 4, the constraint on $b$ for GRB+Planck is only due to GRB and the constraint on $b$ for GRB+Pantheon-Plus is due to both Pantheon-plus and GRB data and the shift in in $b$ is due to the effect of Pantheon-plus data on GRB.

\section{Conclusions}
\label{sec:conclusions}
In summary, our study investigated how Hubble tension affects observations of Gamma-Ray Bursts (GRBs) and Quasi-Stellar Objects (QSO). Referring to the work by Camarena et al. in 2023 [~\cite{Camarena2023}], we found that differences in the absolute magnitude ($M$) of Type Ia supernovae within the Pantheon-Plus sample are influenced by priors from Planck and SH0ES. We calibrated $M$ using the background $\Lambda$CDM model through the likelihood analysis. In Section~\ref{subsec:Mb}, we discussed how these priors impacted the value of $M$. Notably, the correlation between $M$ and $H_0$ became evident, with a decrease in the value of $H_0$ from the Planck prior leading to an increase in $M$, and an increase from the SH0ES prior resulting in a decrease in $M$.

The Pantheon-Plus Sample includes observations in the redshift range of $0.01<z<2.26$. Expanding our investigation, we extended the analysis to redshift $z=5$ for GRBs and even further for QSOs, with observations reaching up to $z=7.5$. Notably, a similar trend in the intrinsic parameters of GRBs was identified throughout our extended analysis. Our initial focus was on examining whether the inclusion of GRB and QSO observations in Pantheon-Plus measurements affected the overall cosmological parameters. Surprisingly, we found no significant alterations in the cosmological parameters obtained from the Planck likelihood when integrating GRB, QSO, and Pantheon-Plus observations. Following this, we conducted a detailed exploration of the intrinsic parameters of GRBs, considering both early and late-time observations. A noteworthy finding was a substantial discrepancies in intrinsic parameters from local and high-redshift observations. Given the GRB observations extending to higher redshifts, we initially considered the possibility of redshift evolution, as suggested in the work by Dainotti et al.~\cite{Dainotti2023}, to comprehend the implications of this discrepancy. Incorporating the redshift evolution corrections substantially effected the discrepancies. We found out 
 and showed in the table \ref{tab:all_tab} that in case of GRB corrections with Planck prior, the shift in the GRB intrinsic parameters are not that significant. But we found a significant change for the GRB corrections with Pantheon-Plus\&SH0ES prior. This lead to a substantial reduction in the $b$ GRB intrinsic parameter tension as shown in table \ref{tab:all_tab}. We further expanded our analysis concerning evolutionary corrections. By applying Gaussian priors to the parameters of the evolution correction as described in Equation~\ref{eq:correction}, we notably alleviated tension in the intrinsic parameters. Consequently, the tension in the $b$ parameter decreased to $2.26\sigma$, indicating the significant impact of evolution correction on GRB observations. However, still an indication of this tension remains in the $b$ intrinsic parameter of GRBs, posing a new puzzle for investigation, especially in view of new observations. 
 We also discussed in subsection \ref{eq:correction} that these discrepancies could be attributed to the conflicts in Pantheon-Plus and GRB data measurements of luminosity distance at certain redshift values. To fully solve this issue one would need to have a much larger sample of GRBs, for reaching a much higher precision on cosmology \cite{2022MNRAS.514.1828D}  with the aid of machine learning analysis \cite{2024ApJS..271...22D,2024arXiv240204551D} and lightcurve reconstruction\cite{2023ApJS..267...42D}. In future it would be also really interesting to explore these discrepancies in backdrop of different dark energy models. %This would reveal whether these discrepancies are model dependent. 

\section*{Acknowledgments}
AAS acknowledges the funding from SERB, Govt of India under the research grant no: CRG/2020/004347. AAS and SAA acknowledges the use of High Performance Computing facility Pegasus at IUCAA, Pune, India.

\bibliographystyle{JHEP}
\bibliography{grb}

\end{document}